\documentclass[
 reprint,
 amsmath,amssymb,
 aps,
]{revtex4-2}
\usepackage[T1]{fontenc}
\usepackage{graphicx}
\usepackage{dcolumn}
\usepackage{bm}
\usepackage{amsmath}
\usepackage{lmodern}
\usepackage{mathtools}
\usepackage{float}
\usepackage{ragged2e}

\usepackage{siunitx}
\usepackage[x11names]{xcolor}
\definecolor{LinkColor}{RGB}{46,48,146}
\usepackage[colorlinks,citecolor=LinkColor,linkcolor=LinkColor,urlcolor=LinkColor]{hyperref}
\usepackage[format=plain,labelfont=bf,labelsep=space]{caption}
\DeclareCaptionJustification{mym}{\justifying}
\usepackage[normalem]{ulem}

\setcitestyle{super}
\usepackage[normalem]{ulem}

\usepackage{titlesec}
\titleformat{\section}[display]{\bf\large}{}{0pt}{}
\titleformat{\subsection}[display]{\bf}{}{0pt}{}
\titlespacing\section{0pt}{12pt plus 4pt minus 2pt}{0pt plus 2pt minus 2pt}
\titlespacing\subsection{0pt}{12pt plus 4pt minus 2pt}{0pt plus 2pt minus 2pt}
\titlespacing\subsubsection{0pt}{12pt plus 4pt minus 2pt}{0pt plus 2pt minus 2pt}

\begin{document}

\title{Magneto-optical induced supermode switching in quantum fluids of light}%

\author{Magdalena Furman$^1$}
\author{Rafa{\l} Mirek$^1$}
\author{Mateusz Kr\'ol$^1$}
\author{Wojciech~Pacuski$^1$}
\author{Helgi Sigur{\dh}sson$^{1,2}$}
\author{Jacek Szczytko$^1$}
\author{Barbara Pi\k{e}tka$^1$}
\email{Barbara.Pietka@fuw.edu.pl}

\affiliation{$^1$Institute of Experimental Physics, Faculty of Physics, University of Warsaw, ul.~Pasteura 5, PL-02-093 Warsaw, Poland}
\affiliation{$^2$Science Institute, University of Iceland, Dunhagi 3, IS-107, Reykjavik, Iceland}

\begin{abstract}
The insensitivity of photons towards external magnetic fields forms one of the hardest barriers against efficient magneto-optical control, aiming at modulating the polarization state of light. However, there is even scarcer evidence of magneto-optical effects that can spatially modulate light. Here, we demonstrate the latter by exploiting strongly coupled states of semimagnetic matter and light in planar semiconductor microcavities. We nonresonantly excite two spatially adjacent exciton-polariton condensates which, through inherent ballistic near field coupling mechanism, spontaneously synchronise into a dissipative quantum fluidic supermode of definite parity. Applying a magnetic field along the optical axis, we continuously adjust the light-matter composition of the condensate exciton-polaritons, inducing a supermode switch into a higher order mode of opposite parity. Our findings set the ground towards magnetic spatial modulation of nonlinear light.
\end{abstract}

\maketitle

\section{Introduction}
Intricate magneto-optical effects decorate the optical response of magnetic materials with far reaching applications across sensing, material characterisation, and light modulation~\cite{Lan_NanoSelect2020,Kimel_JPDAP2022}. Such effects can be enhanced in optical structures displaying strong light-matter interactions~\cite{Kavokin_PRB1997} opening new ways in magnetic control over light underpinned by adventures into atomically thin materials~\cite{Lyons_NatPho2022}, hyperbolic metamaterials~\cite{Li_SciAdv2022}, plasmonics~\cite{Armelles_NJP2014}, and more. In particular, dilute magnetic semiconductors~\cite{Diet_NatMatl2010} embedded in planar microcavities (i.e., semimagnetic cavities)~\cite{Rousset_APL2015} host exciton resonances that can be vividly affected by an external magnetic field, in contrast to the bare cavity photons. In the strong coupling regime, these cavities display large energy splitting between the emergent spin-up and spin-down exciton-polariton modes~\cite{Mirek_PRB2017, Rousset_PRB2017, Krol_SciRep2018, Krol_PRB2019, Mirek_PRB2023} due to the underlying electron and hole Zeeman effect.

Exciton-polaritons (from here on {\it polariton}) are bosonic quasiparticles formed by the strong coupling between a quantum well exciton and a cavity photon~\cite{Deng_RMP2010}. Polaritons posses a two-component integer spin structure $s_z = \pm 1$ along the cavity growth axis corresponding to bright excitons excited by right and left-hand circularly polarised photons $\sigma^{\pm}$. Through strong exciton-mediated Coulomb interactions and light effective photon mass, polaritons can undergo a nonequilibrium analogue of Bose-Einstein condensation at elevated temperatures~\cite{Kasprzak2006, Deng_RMP2010}. A salient feature of polaritons quantum fluids is that information on the macroscopic spinor wave function (spin, phase, amplitude, correlations) is encoded in the emitted cavity light subject to standard optical measurement techniques. Condensates of polaritons thus form a coherent source of nonlinear light which can be manipulated using electric fields~\cite{Schneider_Nature2013,Lim_NatComm2017,Loginov_PRR2020} and magnetic fields~\cite{Tignon_PRL1995, Fisher_PRB1996, Walker_PRL2011, Fischer_PRL2014, Caputo_CommPhys2019}. This has led to deep exploration into polaritons as fundamentally nonequilibrium spinor quantum fluids~\cite{Yang_AdvQuantMat2022}, potential candidates for future spinoptronic technologies~\cite{Liew_PhysE2011, Sedov_LSAppl2019}, and testbed for spin-orbit coupling phenomena~\cite{Leyder_NatPhys2007, Sala_PRX2015, Lempicka_SciAdv2022}. They have been demonstrated to contain a plethora of nontrivial physical effects including half-vortices~\cite{Lagoudakis_Science2009}, half solitons~\cite{Hivet_NatPhys2012}, polarization pinning~\cite{Kasprzak_PRB2007, Gnusov_PRB2020}, spontaneous spin bifurcations~\cite{Ohadi_PRX2015}, and self-induced Larmor precesssions~\cite{Sigurdsson_PRL2022} just to name a few.

\begin{figure}[t]
	\centering
    \includegraphics{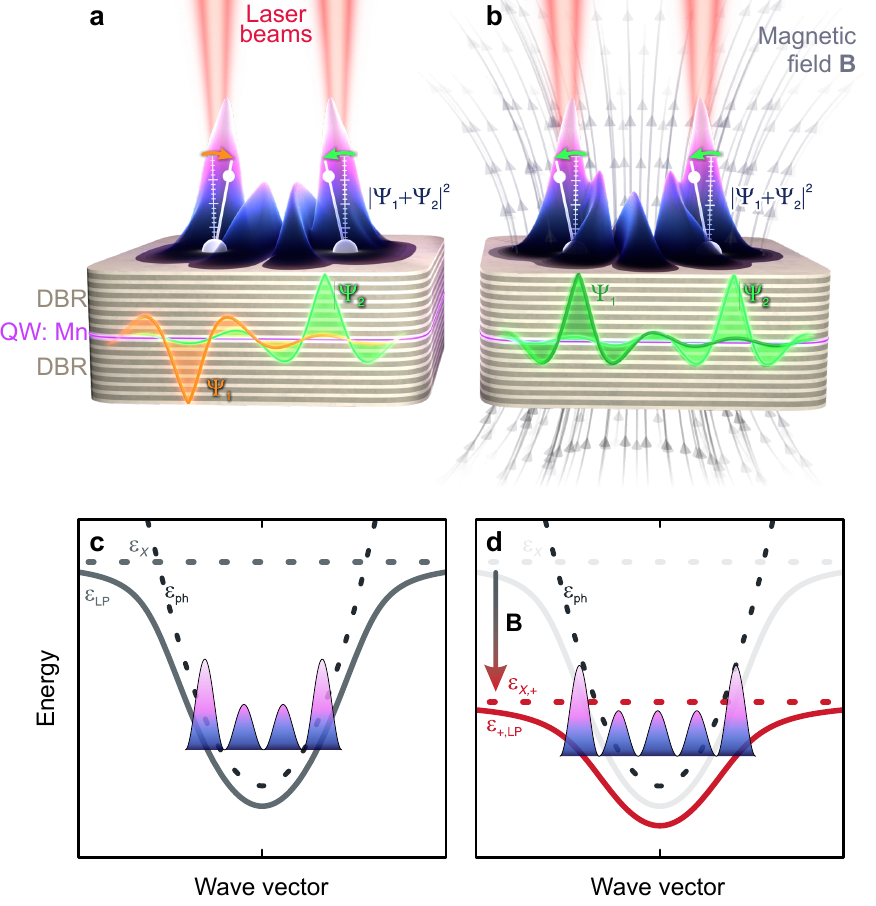}
	\caption{\textbf{Polariton condesate supermode parity switching in magnetic field.} \textbf{a} Schematic image of an odd parity polariton condensate supermode created by a ballistic synchronisation of two condensates excited with two laser beams. \textbf{b}~In external magnetic field the parity changes to even. \textbf{c}~Schematic dispersion relation of the lower polariton mode, with a odd-parity condensate without external magnetic field. \textbf{d} Change of the dispersion relation in magnetic field resulting in the switching of the parity of the condensate to even.}
	\label{im:idea}
\end{figure}

In this study, we utilise semimagnetic cavities to explore an unconventional magneto-optical effect seemingly unique to strongly coupled states of light and matter. We report on the observation of magneto-optical induced transverse spatial mode-switching (synchronization flipping) between two in-plane coupled polariton condensates, a polariton {\it dyad}~\cite{Topfer_CommPhys2020}, as a function of external magnetic field strength despite the polariton's near charge neutrality (see Fig.~\ref{im:idea}a,b for schematic). To achieve this, we utilize a incoherent optical excitation beam structured into two Gaussian pump spots that are focused onto the plane of a semimagnetic optical cavity. Each spot locally excites a polariton condensate characterized by a strong ballistic outflow of coherent particles. The subsequent intricate near field interference between the condensates causes them to phase synchronize into a coherent extended supermode of definite parity~\cite{Topfer_CommPhys2020}. The term supermode refers to modes arising from coupling between degenerate counterpropagating waves~\cite{Kippenberg_OptLett2002} which are represented here by outflowing polaritons. By applying an external out-of-plane magnetic field we induce giant Zeeman splitting in the exciton wavefunction which, through strong coupling, carries into the polariton wavefunction~\cite{Mirek_PRB2017, Krol_SciRep2018, Mirek_PRB2023}. Consequently, the polariton dispersion becomes dramatically reshaped, splitting the effective masses of spin-up and spin-down polaritons and changing the interparticle interaction strength of excitons and polaritons which affects the dyad's coupling mechanism. By gradually increasing the strength of the magnetic field we observe the condensate dyad losing synchronicity, associated with a lack of interference fringe contrast in the cavity near field emission. At even higher field strengths we observe a restoration of the dyad synchronicity into a supermode of opposite parity, evidencing a magnetically induced polariton mode-switching. Our findings are accurately produced through polariton mean field models accounting for the changing exciton wavefunction.

\section{Theory}
\subsection{Magnetically changing dispersion}
In this section, we explain how an incident magnetic field affects the dispersion and light-matter composition of polaritons. The result of which is a strong modification in the ability of two spatially separated condensates to couple and synchronise. We start by considering only the lower polariton energy branch for each spin component $\psi_\pm$ defined by the dispersion relation~\cite{Deng_RMP2010}:
\begin{equation} \label{disp1}
\epsilon_{\pm}(\mathbf{k}) = \frac{\epsilon_\text{ph}(\mathbf{k}) + \epsilon_{X,\pm}}{2} - \sqrt{ \Omega_R^2 + \left( \frac{\epsilon_\text{ph}(\mathbf{k}) - \epsilon_{X,\pm}}{2}\right)^2 }.
\end{equation}
Here, $\Omega_R$ is the light-matter Rabi energy, $\epsilon_\text{ph}(\mathbf{k}) = \hbar^2 k^2 / 2m$ is the cavity photon dispersion in the parabolic approximation, and $\epsilon_{X,\pm} = \epsilon_0 \mp \mu_B g_\text{eff} B$ is the heavy-hole exciton energy. We neglect mixing with the light-hole excitons which are detuned much higher in energy~\cite{Mirek_PRB2017}. The effective exciton $g$-factor is denoted $g_\text{eff}>0$, $\mu_B$ is the Bohr magneton, and $B$ is the normally incident magnetic field (Faraday geometry) [see Fig.~\ref{im:idea}b]. We note that the diamagnetic shift of the exciton resonance in our cavity sample is much weaker than other energy scales in the dispersion and is therefore omitted~\cite{Mirek_PRB2017}.

The exciton and photon Hopfield fractions of the lower polariton spin states follow from standard expressions,
\begin{align} \label{hopfield}
\begin{split}
    |X_\pm|^2 & = \frac{1}{2}\left( 1 - \frac{\epsilon_{X,\pm} - \epsilon_\text{ph}(\mathbf{k})}{\sqrt{ [\epsilon_{X,\pm} - \epsilon_\text{ph}(\mathbf{k})]^2 +  4\Omega_R^2}} \right),  \\ 
    |C_\pm|^2 & = \frac{1}{2}\left( 1 + \frac{\epsilon_{X,\pm} - \epsilon_\text{ph}(\mathbf{k})}{\sqrt{ [\epsilon_{X,\pm} - \epsilon_\text{ph}(\mathbf{k})]^2 +  4\Omega_R^2}} \right). 
    \end{split}
\end{align}
Around small momenta, the energy difference between the two branches $\Delta_Z(\mathbf{k}) = \epsilon_{-}(\mathbf{k}) - \epsilon_{+}(\mathbf{k})$ can be written,
\begin{align} \notag 
\Delta_Z(\mathbf{k}) & \approx 
\mu_B g_\text{eff} B \left(1 - \frac{ \epsilon_0}{\sqrt{4\Omega_R^2 + \epsilon_0^2 + \mu_B^2 g_\text{eff}^2 B^2}}  \right) \\ \label{eq.zeeman}
 & + \frac{\epsilon_\text{ph}(\mathbf{k})}{2}\left[ \frac{\epsilon_{X,-}}{\sqrt{ 4\Omega_R^2 + \epsilon_{X,-}^2 }} - \frac{\epsilon_{X,+}}{\sqrt{ 4\Omega_R^2 + \epsilon_{X,+}^2 }} \right].
\end{align}
The curved bracket describes the polariton Zeeman splitting at $k=0$ to the leading order in light-matter composition determined by the detuning $-\epsilon_0$. The squared brackets describe the leading order correction to the splitting as a function of momentum, showing a quadratic dependence capturing the growing exciton fraction with higher momenta. The momentum dependence of the splitting implies that the masses between spin-up and spin-down have become different under a magnetic field following,
\begin{equation} \label{eq.mass}
m_{\pm} \simeq \frac{m}{|C_\pm|^2}. 
\end{equation}
The result is a significant change in the dispersion of each polariton spin when a magnetic field is turned on as shown schematically in Fig.~\ref{im:idea}c,d. Moreover, for $B>0$ the detuning between spin-up excitons and photons decreases leading to higher matter component $|X_+|^2$ for the spin-up polaritons. This in-turn increases the interparticle interaction strength of spin-up polaritons since $\alpha_\sigma = g_0 |X_\sigma|^4$ where $g_0$ is the bare exciton-exciton Coulomb interaction strength. This magnetic sensitivity of the spin-anisotropic particle interaction carries directly into the coupling mechanism of the two condensates as we will see.

\subsection{Condition of magnetically induced supermode switching}
Recently, optical write-in techniques to create reprogrammable structured polariton condensates made some serious advances, achieving hundreds of coherently interacting condensates across various graph topologies~\cite{topfer2021engineering, Pickup2020}. Tightly focused excitation (pump) spots lead to strong local blueshifts for polaritons while also providing optical gain, resulting in a highly nonequilibrium condensate characterised by a coherent outflow of polaritons with high velocities that could interact with distant neighbours~\cite{Alyatkin_PRL2020}. Unlike low-momentum condensates, these interacting ballistic condensates display extremely intricate interference fringes (much smaller than the condensate healing length), regimes of robust synchronisation bridged by limit cycle behaviours~\cite{Topfer_CommPhys2020}, and perseverance of spin information over macroscopic distances~\cite{Kammann2012, Anton_PRB2015}. The coupling between adjacent polariton condensates can be adjusted through external parameters such as pump power and distance~\cite{Ohadi_PRX2016}, optical barriers~\cite{Alyatkin_PRL2020}, and exciton-photon detuning. However, magnetic control over the coupling has not been possible to date.

The coupling between adjacent ballistic condensates depends on their relative distance which defines the lossy supermode they synchronise to form~\cite{Ohadi_PRX2016, Topfer_CommPhys2020}. This corresponds to the distance between the lasers as shown in Fig.~\ref{im:idea}a,b. Outflowing polaritons take time to reach their neighbouring condensate which results in accumulation of phase as they propagate,

\begin{equation}
    \Delta \phi \approx k_c d. 
\end{equation}
Here, $k_c$ is the average outflow wavevector of each condensate and $d$ is their separation distance. If $\Delta \phi = \pi \ (\text{mod} \ 2\pi)$ then the condensates synchronize into an anti-phase supermode of odd parity, whereas if  $\Delta \phi = 0 \ (\text{mod} \ 2\pi)$ they form an in-phase supermode of even parity (see Fig.~\ref{im:idea}a,b). An applied magnetic field in our sample is able to change $\Delta \phi$ by an amount $\pi$, triggering a switch between modes.

The wavevector $k_c$ can be estimated by considering the elastic conversion of the reservoir-induced potential energy into polariton kinetic energy (see Eq.~\eqref{kcB2} in Methods)
\begin{equation} \label{kcB}
    k_{c,\sigma}  \approx  \sqrt{\frac{4g_0m}{N_\text{QW}\hbar^2 } \frac{|X_\sigma|^2}{|C_\sigma|^2} n_{X,\sigma}},
\end{equation}
where $n_{X,\sigma}$ is the density of reservoir excitons (see Eq.~\eqref{eq.nX} in Methods), and $N_\text{QW}$ is the number of quantum wells. The presence of the exciton Hopfield fraction $|X_\sigma|^2$ in the above signifies the importance of the magnetic sensitivity of the spin-anisotropic polariton interactions. We have assumed that the blueshift dominantly comes from the reservoir and not the condensate density, which is a safe assumption when working not too far above the condensation threshold.

\begin{figure*}
\centering
\includegraphics[width=\linewidth]{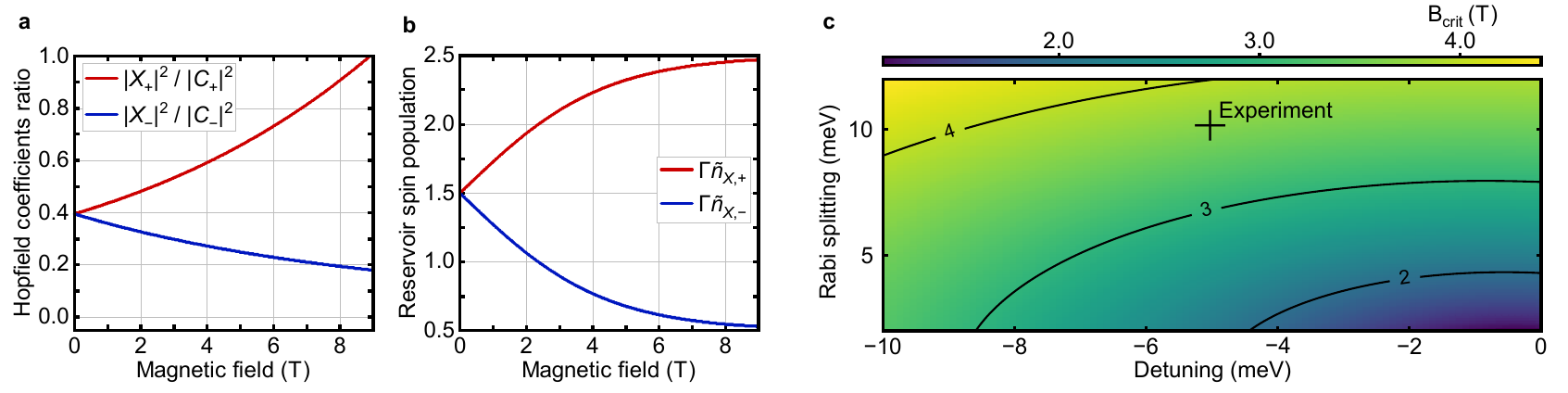}
\caption{\textbf{Change of polariton parameters as a function of increasing magnetic field.} 
\textbf{a,b} Dependence of the \textbf{a} polariton light-matter composition~\eqref{hopfield} at $k=0$ and the \textbf{b} reservoir spin populations on magnetic field~\eqref{eq.nX}. \textbf{c}~Calculated critical magnetic field strength $B_\text{crit}$~\eqref{Bcrit} needed to induce mode switching in the polariton dyad.}
\label{theory1}
\end{figure*}

The change in the momentum $k_{c,\sigma}$ of ballistically propagating polaritons with magnetic field can be better understood by looking separately at the contribution coming from the Hopfield fractions and the reservoir. Figure~\ref{theory1}a,b shows the change in the polariton light-matter composition $|X_\sigma|^2 / |C_\sigma|^2$ (at $k=0$) and the reservoir spin populations for parameters corresponding to our sample~\cite{parameters}, as function of magnetic field, respectively. For positive magnetic fields, the exciton fraction of spin-up polaritons increases as the detuning between the exciton level and photon closes (see Fig.~\ref{im:idea}d for schematic). Spin-down excitons, however, move away from the photon branch, reducing the exciton Hopfield fraction of polaritons. For this reason, Figure~\ref{theory1}a shows opposite trends in the ratios of the Hopfield fractions for each spin. Figure~\ref{theory1}b shows an increase in the spin-up reservoir population with positive $B$ because of their slower spin relaxation rate [see Eq.~\eqref{eq.spin_relax} in Methods]. 

Combined, these effects serve to increase $k_{c,+}$ (for $B>0$) according to~\eqref{kcB}. This eventually triggers a mode switching event in the polariton dyad with loss of synchronicity. For a given power, this magnetically induced mode switching occurs when,
\begin{equation}
    \pi = |k_{c,\sigma}(B_\text{crit}) - k_{c,\sigma}(0) | d.
\end{equation}
Assuming the condensate supermode possesses a characteristic harmonic wavelength defined by $\lambda_c = 2\pi/k_c \approx 2 d/n$ where $n \in \mathbb{N}$ we obtain an equation defining the critical magnetic field strength,
\begin{equation} \label{Bcrit}
   % \left( \frac{\pi}{d}(1+n) \right)^2 = \frac{4g_0m}{N_\text{QW}\hbar^2 } \frac{|X_\sigma|^2}{|C_\sigma|^2}  P_0 \tilde{n}_{X,\sigma}. 
      \left( \frac{1+n}{n} \right)^2 \left( \dfrac{|X_\sigma|^2}{|C_\sigma|^2}   \tilde{n}_{X,\sigma} \right)_{B=0} = \left( \dfrac{|X_\sigma|^2}{|C_\sigma|^2}   \tilde{n}_{X,\sigma} \right)_{B=B_\text{crit}}. 
\end{equation}
Figure~\ref{theory1}c shows the $B_\text{crit}$ solutions to Eq.~\eqref{Bcrit} as a function of Rabi splitting and detuning with the black cross denoting the current experiment. The results predict that mode-switching should be possible within only a few Tesla. We note that the Hopfield coefficients are evaluated at $k=0$ since polaritons condensing on top of the pump spots are low momentum.

\subsection{Numerical results}
\begin{figure}
\centering
\includegraphics[width=\linewidth]{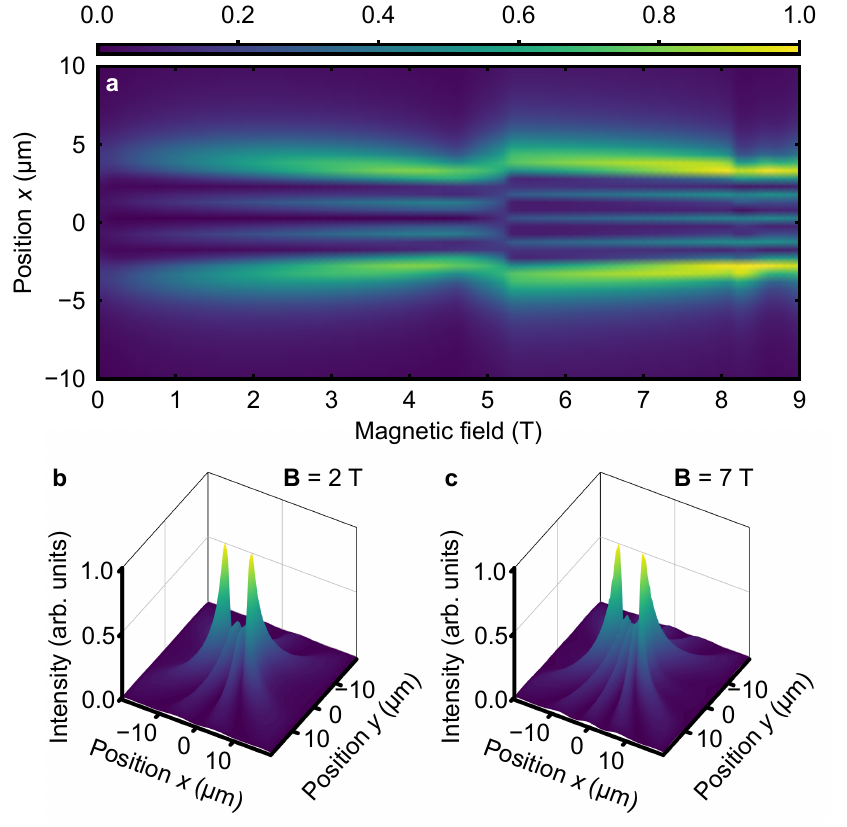}
\caption{\textbf{Theoretical simulation showing change of supermode parity as a function of the increasing magnetic field.} \textbf{a} Time integrated line profile (i.e., along $y=0$ direction) of the total condensate density as a function of magnetic field. \textbf{b,c} Corresponding $x$-$y$ density profiles at low and high magnetic fields.}
\label{theory2}
\end{figure}

To test the feasibility of the magnetically induced mode-switching we numerically solve the two-dimensional spinor generalized Gross-Pitaevskii model, describing the condensate order parameter $\Psi(\mathbf{r},t) = (\psi_+, \psi_-)^\text{T}$, coupled to an exciton reservoir rate equation [see Eqs.~\eqref{eq.GPE} and~\eqref{eq.Res} in Methods]. We use random white noise initial conditions and damped boundary conditions. The results are shown in Fig.~\ref{theory2}a where we plot the normalized time-averaged line profile (i.e., along $y=0$ direction) of the total condensate density $\langle \Psi^\dagger \Psi \rangle$ with varying magnetic field. For each value of $B$ we use a different independent white noise initial condition. The simulated time window for each value of $B$ is 3 ns or almost $10^3$ longer than the polariton lifetime. 

\begin{figure*}
\centering
\includegraphics[width=\linewidth]{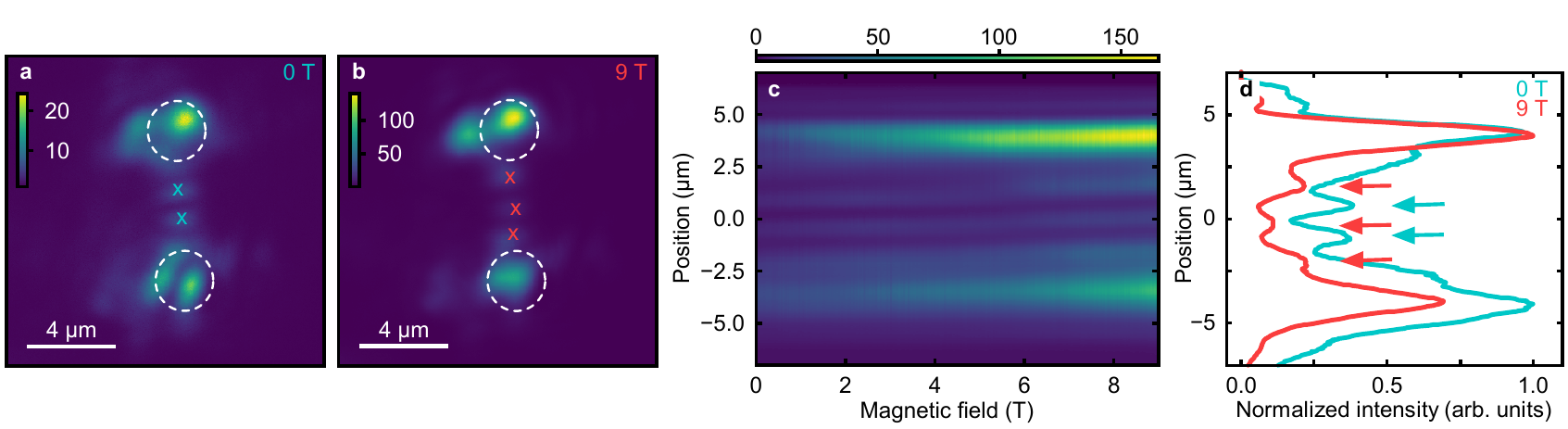}
\caption{\textbf{Real space condensate supermode parity change as a function of increasing magnetic field.} Emission from the condensate supermode \textbf{a} without external magnetic field and \textbf{b} at magnetic field of 9\,T. %Positions of constructive interference fringes are marked with blue circles and red crosses. 
\textbf{c}~Emission intensity profile along the line connecting the %excitation sites
condensates and its evolution with the applied external magnetic field. \textbf{d} Cross sections of \textbf{c}~at 0\,T and 9\,T magnetic fields. Arrows marks positions of the constructive interference fringes.}
\label{im:exp1}
\end{figure*}

At low magnetic fields, the system condenses into a odd-parity solution ($n=3$) and, as magnetic field is increased, experiences a small contraction in its profile (see Fig.~\ref{theory2}a) before switching to the next high-order mode of even-parity ($n=4$). Corresponding $x-y$ spatial maps of the condensate density at low and high magnetic fields are shown in panels Fig.~\ref{theory2}b,c. We note the increasing density of the condensate with magnetic field which reflects the growing spin population imbalance in the reservoir, resulting in higher gain for spin-up polaritons for $B>0$. We also point out that the $B_\text{crit} \approx 5$ T estimated from looking at Fig.~\ref{theory2}a is slightly higher than the predicted value from~\eqref{Bcrit} shown in Fig.~\ref{theory1}c. This slight discrepancy stems from the assumption $\lambda_c \approx 2d/n$ which is only valid for ideal planar Fabry-P\'{e}rot resonators. 

\section{Experimental Results}
To confirm our theoretical predictions, we conducted an experiment on a semimagnetic microcavity made of II--VI semiconductors~\cite{Rousset_APL2015} subjected to an external magnetic field. The cavity structure is composed of (Cd, Zn, Mg)Te and contains six quantum wells doped with 0.5$\%$ manganese ions resulting in giant Zeeman splitting of the polariton spins~\cite{Mirek_PRB2017,Rousset_PRB2017,Krol_SciRep2018, Krol_PRB2019, Mirek_PRB2023}. For details about the sample see Methods and Ref.\,\cite{Mirek_NanoLett2021}.  All measurements were carried out at a temperature of 5~K. We excited the sample nonresonantly with a linearly polarized tunable pulsed laser (pulse width $\approx 3$~ps). The laser was tuned to the first high energy Bragg minimum of the cavity. The excitation was split into two Gaussian beams (FWHM $\approx 2$~$\mu$m) incident on the sample at the same time, focused on the surface of the cavity and separated by $d \approx 8$~$\mu$m. The intensity of both laser beams is the same and is set above the condensation threshold ($P_\text{exc}\approx 1.5 P_\text{th}$). The emission from the microcavity was detected using a longpass filter to cut out the reflected laser.

Figure~\ref{im:exp1}a shows the emission from the sample without external magnetic field. We observe that the PL is spatially extended in the whole region between the two excitation spots (marked with grey dashed circles). The emission shows that the condensate at each pump spot has synchronized and formed a spatial supermode (i.e., polariton dyad), evidenced by the appearance of the interference fringes between the excitation sites. The odd parity of the condensate supermode can be observed by the presence of two constructive interference fringes, marked in Fig.~\ref{im:exp1}a by blue crosses. We note that our sample contains large levels of disorder which would normally inhibit coherent coupling between the two condensates. Therefore, the observation of coherent coupling between the two condensate nodes underpins the efficiency of the ballistic coupling mechanism.

\begin{figure*}
\centering
\includegraphics[width=\linewidth]{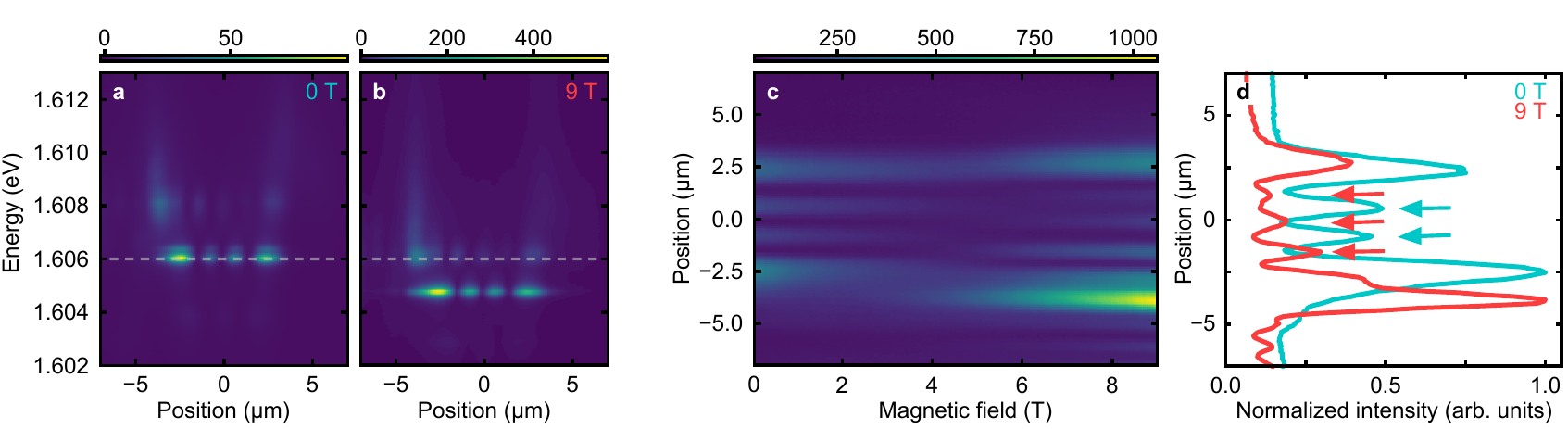}
\caption{\textbf{Energy-resolved condensate supermode parity change as a function of increasing magnetic field.} Spatially-resolved emission spectra from the condensate supermode: \textbf{a} without external magnetic field and \textbf{b}~at magnetic field of 9\,T. \textbf{c} Cross section at energy 1.606\,eV (marked with dashed grey lines on \textbf{a} and \textbf{b} in external magnetic field. \textbf{d} Cross sections of \textbf{c} at 0\,T and 9\,T magnetic fields. Arrows marks positions of the constructive interference fringes.}
\label{im:exp2}
\end{figure*}

When a 9\,T magnetic field is applied to the sample we observe condensation into a higher order supermode of even parity (see Fig.~\ref{im:exp1}b) with three interference fringes (red crosses). This evidences that the predicted magnetically induced mode-switching has occurred, regardless of the charge neutrality of polaritons. Figure~\ref{im:exp1}c shows the evolution of the condensate dyad cross section as a function of magnetic field going across the mode-switching point. The corresponding line profiles at 0\,T and 9\,T are presented in Fig.~\ref{im:exp1}d showing the two different interference (standing wave) patterns. Because of the inherent noisy-ness of the experiment from e.g. pump fluctuations, reservoir induced quantum noise effects, and sample disorder the transition between modes is smeared out. This is in contrast to the sharper switch observed in simulation (see Fig.~\ref{theory2}a) which assumes an ideal coherent system.

We also resolved the emission in energy by directing the image onto a spectrometer. Fig.~\ref{im:exp2}a presents the resulting spectrum from the condensate supermode PL at 0\,T. We observe two competing condensate modes, one with even parity visible for energy of 1.608\,eV and second with odd parity at 1.606\,eV. When a magnetic field is applied the observed modes decrease monotonically in energy in agreement with the Zeeman effect which redshifts the dominant condensate spin population. We note that in the presence of the $B>0$ magnetic field the emission is highly right-hand circularly polarized~\cite{Mirek_PRB2023} corresponding to a dominant spin-up population in the condensate $|\psi_+|^2\gg|\psi_-|^2$. The vice versa is true for $B<0$.

At 9\,T the odd parity condensate supermode energy shifts to 1.605\,eV and the even parity one to 1.606\,eV, as visible in Fig.~\ref{im:exp2}b. Energy filtering the emission at 1.606\,eV (dashed grey lines in Figs.~\ref{im:exp2}a,b) while scanning the magnetic field we observe a much sharper mode-switching in Fig.~\ref{im:exp2}c in contrast to energy-integrated measurements in Fig.~\ref{im:exp1}. The corresponding line profiles at 0\,T and 9\,T are presented in Fig.~\ref{im:exp2}d. We note that energy resolved measurements reveal a substantial portion of the condensate still resides in the original lower-energy supermode implying the energy relaxation effects are quite strong in our sample as one would expect for a highly interactive system.

\section{Discussion}
In this study, we have presented theoretical and experimental evidence of magnetically induced supermode switching in optically structured exciton-polariton condensates in a semimagnetic II-VI CdTe microcavity doped with manganese atoms. Our findings are in contrast to standard magneto-optical effects that typically aim to modulate the plane of polarisation of incident light with the Faraday and Kerr effects being undoubtedly the most famous examples.

Instead, our idea is based on the unique combination of ballistic exciton-polariton condensates~\cite{Topfer_CommPhys2020, topfer2021engineering} and the large levels of Zeeman splitting obtainable between spin-up and spin-down bright semimagnetic excitons. We exploit the strong light-matter coupling between semimagnetic excitons and cavity photons to create extended polariton quantum fluids whose transverse spatial degrees of freedom---as well as polarisation---can be modulated with applied normally incident magnetic field. Our findings are reminiscent of strong spin-orbit coupling effects but are instead entirely based on the light-matter composition of the excited condensate polaritons. In particular, we show that the spontaneous synchronisation between two adjacent ballistic condensates can be flipped from in-phase to anti-phase within a few Tesla. In other words, we demonstrate magnetically induced flip in the parity of the condensate supermode. Our findings do not rely on any polarimetry techniques as the switching event can be fully detected in the near-field and far-field photoluminescence structure of the total emitted cavity light. 

We believe that magnetic control over the transverse spatial degrees of freedom in nonlinear light offers exciting perspectives in shaping and exploring nontrivial phases within large-scale programmable polariton networks~\cite{topfer2021engineering} and analogue simulation of XY systems~\cite{Berloff2017, Tao_NatComm2022}. Previous techniques on controlling the coupling between condensates rely on changing the power and reshaping the nonresonant excitation beam using spatial light modulators~\cite{topfer2021engineering}, introducing auxiliary optical barrier beams~\cite{Alyatkin_PRL2020}, or effective spin-orbit coupling to screen interactions~\cite{Aristov_PRB2022, dovzhenko2023nearest} Here, we offer a solution to adjust the coupling strength in polariton networks without changing the parameters of the pump. 

Moreover, excitons by themselves play an important role in spintronics and optoelectronics, implying that exciton-polaritons could play an important role in future spinoptronic applications with magnetic control~\cite{Liew_PhysE2011}. Finally, strong magneto-optical effects are essential in topological photonics~\cite{Lu_NatPho2014} and polaritonics~\cite{Solnyshkov_OptMatExp2021} for robust lasing and symmetry protected signal transmission. The effect reported here could open new avenues within topological polaritonics when combined with more complicated structured optical beams, or patterned cavities containing large effective photonic spin-orbit coupling~\cite{Rechcinska_Science2019}. 

\section{Methods}

\subsection{Semimagnetic microcavity}
The investigated semiconductor optical microcavity was grown using molecular beam epitaxy on a (100)-oriented GaAs substrate. It was made up of two distributed Bragg reflectors (DBRs), which consisted of alternating Cd\textsubscript{0.874}Zn\textsubscript{0.033}Mg\textsubscript{0.093}Te and Cd\textsubscript{0.35}Mg\textsubscript{0.65}Te layers. The top DBR consisted of 16 layer pairs, while the bottom one of 19 pairs. The cavity between these two DBRs was formed by approx. 600\,nm thick Cd\textsubscript{0.874}Zn\textsubscript{0.033}Mg\textsubscript{0.093}Te containing three pairs of 20\,nm thick Cd\textsubscript{0.962}Zn\textsubscript{0.033}Mn\textsubscript{0.005}Te quantum wells. The low concentration of magnetic manganese ions, approximately 0.5$\%$, is present only in the quantum wells. The quality factor of the microcavity at 5\,K based on an angle-resolved reflection measurement was around 300.

\subsection{Theory}
Here we detail on the two-dimensional spatiotemporal modelling of the magnetic polariton condensate. Polariton condensation can be accurately modelled in the mean field picture using generalised Gross-Pitaevskii equations coupled to exciton reservoir rate equations. The condensate spinor order parameter is denoted $\Psi(\mathbf{r},t) = (\psi_+, \psi_-)^\text{T}$ and the active exciton reservoir $n_{A,\pm}(\mathbf{r},t)$,

\begin{align} \notag
     i\hbar \frac{\partial \psi_{\pm}}{\partial t}  &= \bigg[\epsilon_{\pm}(-i\boldsymbol{\nabla}) + \frac{i \hbar R n_{A,\pm} }{2} \\
    & + \alpha_\pm |\psi_{\pm}|^{2}  + G_\pm \left(n_{A,\pm} + n_{I,\pm} \right) \bigg]\psi_{\pm} ,
    \label{eq.GPE} \\ \notag
     \frac{\partial n_{A,\pm}}{\partial t}  &= -\left(\Gamma  + \Gamma_{s,\pm} + R|\psi_{\pm}|^2\right)n_{A,\pm} \\
     & + \Gamma_{s,\mp} n_{A,\mp} +  W n_{I,\pm}(\mathbf{r}).
    \label{eq.Res}
\end{align}

Here, $\epsilon_{\pm}(-i\boldsymbol{\nabla})$ is adjusted to account for the cavity losses by the replacement $\epsilon_\text{ph}(-i\boldsymbol{\nabla}) \to \epsilon_\text{ph}(-i\boldsymbol{\nabla}) - i \hbar/2\tau_\text{ph}$ where $\tau_\text{ph}$ is the photon lifetime. Other parameters are: $G_\pm = 2 g_0 |X_\pm|^2/N_\text{QW}$ and $\alpha_\pm = g_0 |X_\pm|^4/N_\text{QW}$ are the same-spin polariton-reservoir and polariton-polariton interaction strengths normalized over the number of quantum wells, respectively, $R$ is the scattering rate of reservoir excitons into the condensate, $\Gamma$ is the active reservoir decay rate, $\Gamma_{s,\pm}$ describes the rate of spin relaxation for each spin component, $W$ quantifies the conversion of dark and high-momentum inactive excitons $n_{I,\pm}$ into the active $n_{A,\pm}$ exciton reservoir.  We neglect the opposite-spin interactions since they are much weaker than the same-spin interactions~\cite{Bieganska_PRL2021}. All parameter values can be found in~\cite{parameters}.

\begin{figure}
\centering
\includegraphics[width=\linewidth]{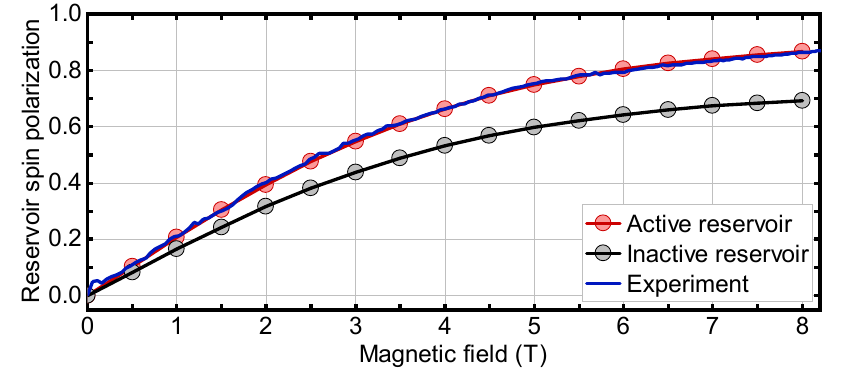}
\caption{\textbf{Reservoir spin polarization.} Normalised spin polarisation in the steady state solutions of the active Eq.~\eqref{eq.Res} and inactive reservoir Eq.~\eqref{eq.ResIn} below threshold. Equation~\eqref{eq.spin_relax} is used to fit the solution of Eq.~\eqref{eq.Res} to the experimentally measured circular polarisation in the cavity emission.}
\label{theory3}
\end{figure}

In the steady state the inactive reservoir, driven by a continuous wave pump, can be estimated as~\cite{Pickup2021}:
\begin{equation}
W\begin{pmatrix} 
n_{I,+} \\
n_{I,-}
\end{pmatrix} \label{eq.ResIn} = \frac{P_0 f(\mathbf{r})}{W+\Gamma_{s,+} +\Gamma_{s,-} } \begin{pmatrix}
W \cos^2{(\theta)} + \Gamma_{s,-} \\ W \sin^2{(\theta)} + \Gamma_{s,+} \end{pmatrix}.
\end{equation}
Here, $P_0$ denotes the power density of the pump which, in this experiment, has a spatial profile $f(\mathbf{r}) = e^{-a|\mathbf{r}-\mathbf{d}/2|^2} + e^{-a|\mathbf{r}+\mathbf{d}/2|^2}$, and $\theta$ defines the ellipticity of the incident excitation.

A unique feature of our model, is that we include the dependence of the exciton spin relaxation parameters $\Gamma_{s,\pm}(B)$ on the magnetic field~\cite{Crooker_PRB1997} which can be heuristically modelled as,
\begin{equation} \label{eq.spin_relax}
\Gamma_{s,\pm} = \Gamma_s \mp \gamma_s \tanh{(B/B_0)}.
\end{equation}
Physically, the magnetic field splits the spin relaxation rates of the excitons leading to a population imbalance in the reservoir spins resulting in finite circular polarisation in the below-threshold cavity emission~\cite{Krol_SciRep2018}. Here, $\Gamma_s$ is the average spin relaxation rate of excitons without an external magnetic field~\cite{Maialle_PRB1993} and $\gamma_s$ and $B_0$ are fitting parameters. The values of these parameters can be established by by fitting the stead-state spin polarization of the active reservoir to the circular polarization in the below-threshold emission of the cavity (see Fig.~\ref{theory3}).

With $\gamma_s$ and $B_0$ established, they define the amount of spin imbalance in the reservoirs corresponding to changing blueshifts for spin-up and spin-down polaritons with magnetic field. Note, this blueshift is dictated by the last term on the RHS in Eq.~\eqref{eq.GPE} which contributes towards the mode switching in the condensate dyad. Equation~\eqref{kcB} is obtained through these terms as,
\begin{equation} \label{kcB2}
k_{c,\sigma} \approx \frac{2m_\sigma}{\hbar^2}  G_\sigma  (n_{A,\sigma} + n_{I,\sigma}) = \frac{4g_0m}{N_\text{QW}\hbar^2 } \frac{|X_\sigma|^2}{|C_\sigma|^2} n_{X,\sigma}
\end{equation}
where, for $\theta=45^\circ$ (linearly polarized excitation), the steady state solution of the combined reservoirs close to threshold  ($|\psi_\sigma|^2 \simeq 0$) is,
\begin{align} \label{eq.nX}
\begin{split}
    n_{X,\sigma} &  = n_{A,\sigma} + n_{I,\sigma} = \frac{P_0}{W+2\Gamma_s} \bigg[ \frac{1}{2} + \frac{\Gamma_{s,-\sigma}}{W} + \\ 
    & \frac{(\Gamma + \Gamma_{s,-\sigma})(\frac{W}{2}+\Gamma_{s,-\sigma}) + \Gamma_{s,-\sigma}(\frac{W}{2}+\Gamma_{s,\sigma})}{\Gamma(\Gamma+2\Gamma_s)} \bigg].
\end{split}
\end{align}

\bibliography{biblio}

\section*{Data availability}
\noindent All data that supports the conclusions of this study are included in the article. The data presented in this study is available from the corresponding author upon reasonable request.

\section*{Acknowledgments} 
\noindent This work was supported by the National Science Center, Poland, under the projects 2020/37/B/ST3/01657 (M.~F., B.~P.) and 2019/33/N/ST3/02019 (R.~M.), and No. 2022/45/P/ST3/00467 (H.~S.) co-funded by the European Union Framework Programme for Research and Innovation Horizon 2020 under the Marie Skłodowska-Curie grant agreement No. 945339.

\section*{Author contributions}
\noindent H.~S. conceived the idea and developed the theoretical description, M.~F., R.~M. and M.~K. performed the experiments, W.~P. grew the sample. H.~S., M.~F., M.~K. and R.~M. wrote the manuscript with input from all other authors, H.~S., J.~S., and B.~P. supervised the project.

\section*{Competing interests} 
\noindent The authors declare no competing interests.

\section*{Additional information} 
\noindent{\bf Correspondence and requests for materials} should be addressed to B.P.

\end{document}